\documentclass[aps,showpacs,showkeys,superscriptaddress,floatfix,preprint,12pt]{revtex4}
\usepackage{graphicx}
\usepackage{color}
\begin{document}
\title{Pair Creation in QED-Strong Pulsed Laser Fields Interacting with
Electron Beams} 
\author{Igor V. Sokolov}
\email{igorsok@umich.edu}
\affiliation{Space Physics Research Laboratory, University of Michigan, Ann
Arbor, MI 48109 }
\author{Natalia M. Naumova}
\affiliation{Laboratoire d'Optique Appliqu\'{e}e, 
UMR 7639 ENSTA, Ecole Polytechnique, CNRS, 91761 Palaiseau, France}
\author{John A. Nees}
\affiliation{Center for Ultrafast Optical Science and FOCUS Center, University
of Michigan, Ann Arbor, MI 48109}
\author{G\'{e}rard A. Mourou}
\affiliation{Institut de la Lumi\`{e}re Extr\^{e}me,  
UMS 3205 ENSTA, Ecole Polytechnique, CNRS, 91761 Palaiseau, France}
\date{\today}
\begin{abstract}
QED-effects are known to occur in a strong laser pulse
interaction with a counterpropagating electron beam, among these
effects being electron-positron pair creation. We discuss the range
of laser pulse intensities of 
$J\ge 5\cdot 10^{22}\ {\rm W/cm^2}$ combined with electron beam 
energies of tens 
of
GeV. In this regime multiple pairs may be generated from a
single beam electron, some of the newborn particles being capable of
further pair production. Radiation back-reaction 
prevents avalanche development and
limits pair creation. The system of integro-differential kinetic
equations for electrons, positrons and $\gamma$-photons is derived and solved numerically.
\end{abstract}


\maketitle
\section{Introduction}
The effects of quantum electrodynamics (QED) may occur in a strong
laser pulse interaction with a counterpropagating electron beam. 
In the well-known experiment \cite{bb} these effects were weak and 
barely observable. If the laser pulse intensity is increased up to 
$J\ge 5\cdot 10^{22}\ {\rm W/cm^2}$ the QED effects control the
laser-beam interaction and result in multiple pair production from a 
single beam electron. 

{\bf QED strong fields.} In QED an electric field, $E$, should be treated as  
strong if it exceeds the Schwinger limit: $E\ge E_S=
m_ec^2/(|e|\lambdabar_C)$ (see \cite{schw}). Such field is
potentially capable of 
separating a virtual electron-positron pair providing an energy, which exceeds the electron rest 
mass energy, $m_ec^2$, to a charge, $e=-|e|$, over an acceleration length as small 
as the Compton wavelength,
$
\lambdabar_C=\hbar/(m_ec)\approx 3.9\cdot10^{-11}{\rm cm}.
$
Typical effects in QED strong fields are: 
electron-positron pair creation from high-energy photons, high-energy photon 
emission from electrons or positrons and the cascade 
development (see \cite{Mark}- \cite{kb}) resulting from the first two processes. 

Less typical is direct pair
separation from vacuum. This effect may only be significant if the
field invariants as defined in \cite{ll}, $F_1=({\bf B}\cdot{\bf E})$, $F_2=E^2-B^2$,
are large enough. 
Here  
the case of {\it weak} field
invariants is considered: 
$|F_{1,2}|\ll E_S^2$, 
and any corrections of the order of $F_{1,2}/E_S^2$ are
neglected (see \cite{dep} about such corrections). Below, 
the term 'strong field' is only applied to the field {\it experienced by a
charged particle}.
 
{\bf QED-strong laser fields.} 
QED-strong fields may be created in 
the focus of an ultra-bright laser.  
Consider QED-effects in  
a relativistically strong pulsed field \cite{Mark}:  
\begin{equation}\label{eq:strong}
|{\bf a}|\gg1,\qquad{\bf a}=\frac{e{\bf A}}{m_ec^2},
\end{equation} 
${\bf A}$ being the vector potential of the wave.   
In the laboratory frame of reference the electric field is not QED-strong for 
achieved laser intensities, $J\sim10^{22}\ {\rm W/cm^2}$ \cite{1022}, and 
even for the $J\sim10^{25}\ {\rm W/cm^2}$ intensity projected
\cite{ELI}. Moreover, both field invariants vanish for 1D waves,  
reducing
the probability of 
direct pair creation from
vacuum  
by
virtue of the
laser field's 
proximity to 1D wave.

Nonetheless, a counterpropagating particle in a 1D  wave, ${\bf
a}(\xi),\,\xi=\omega t-({\bf k}\cdot{\bf x})$, may experience a QED-strong
field,  
$E_0=|d{\bf A}/d\xi|\omega({\cal E}-p_\|)/c$, because the laser
frequency, 
$\omega=c/\lambdabar$, is Doppler
upshifted in the frame of reference co-moving with the electron. 
Herewith the electron dimensionless energy, 
${\cal E}$, and
its momentum are related to $m_ec^2$, and $m_ec$ correspondingly,
and subscript $\|$ herewith denotes the vector projection on the direction of the
wave propagation.  The Lorentz-transformed field exceeds the Schwinger limit, 
if  
$\chi\sim E_0/E_S\ge1$.  
Numerical values of the parameter, $\chi$, may be
expressed in terms of the local instantaneous intensity of the laser
wave, $J$:
\begin{equation}\label{eq:2} 
\chi=\frac32 
\frac{\lambdabar_C}{\lambdabar}({\cal
  E}-p_\|)\left|\frac{d{\bf a}}{d\xi}\right|
\approx\frac{({\cal
  E}-p_\|)}{1.4\cdot10^3}\sqrt{\frac{J}{10^{23}[{\rm W/cm}^2]}}.
\end{equation}

In the SLAC experiment \cite{bb} an electron beam of 
energy $\approx 46.6$ GeV interacted with a counterpropagating terawatt laser
pulse of intensity $J\sim10^{18}{\rm W/cm}^2$ ($|{\bf a}|\le 1$). A value of $\chi\approx
0.4$ had been achieved. An increase in the  
laser field intensity up to $\sim5\cdot10^{22}{\rm W/cm}^2$ ($|{\bf
  a}|\approx 110$) with the use of the same electron beam, would allow us to reach a regime of multiple pair creation
at $\chi\approx90$.

{\bf Radiation back-reaction.} The creation of pairs in QED-strong
fields is a particular form of the radiation losses from charged
particles. At high $\chi$ an intermediate stage in the pair creation
process is the emanation of a high-energy photon by a charged particle: 
$e\rightarrow \gamma,e$ 
(in contrast with $\chi\le1$ case, in which  the ``equivalent''
photons from the electron Coulomb potential mostly contribute to the
pair creation - see \cite{kb} and the papers cited therein). 
This photon is then absorbed in the strong field,
generating an electron-positron pair:
$\gamma\rightarrow e,p$. 

Although the energy-momentum {\it gain} from the strong
laser field is crucial in the course of both emission and pair
creation, still a way to quantify the irreversible radiation {\it
losses} may be found. Specifically, in the 1D wave field the
transfer of energy, $\Delta{\cal E}$, from the wave to a particle may
be interpreted as the absorption of some number of photons, $n$:
$\Delta{\cal E}=n\hbar\omega/(m_ec^2)$. Accordingly, the momentum from the absorbed
photons is added to the parallel momentum of the particle: 
$\Delta p_\|=n\hbar k/(m_ec)=n\hbar
\omega/(m_ec^2)$. So, both energy and parallel momentum are
not conserved, however, their difference is: $\Delta( {\cal E}-
p_\|)=0$. To get the Lorentz-invariant formulation, introduce the
four-vector of the particle momentum, $p=({\cal E},{\bf p})$, and the
wave four-vector, $k=(\frac\omega{c},{\bf k})$ for the 1D wave field. Their four-dot-product, $(k\cdot p)=\omega({\cal
E}- p_\|)/c$, 
is conserved in any particle interaction with the 1D wave field,
including its motion, photon emission, pair creation etc. The
sum of this quantity, $\sum{(k\cdot p_f)}$, over
all particles in the final ({\it f}) state is equal to that for the particles in the 
initial ({\it i}) state: $\sum{(k\cdot p_f)}=\sum{(k\cdot p_i)}$. Each term in
this conserving sum is positive (we use the metric signature $(+,-,-,-)$). Therefore,
any contribution to this sum from a newborn  particle exacts a contribution from its parent.  

Regarding the high-energy electron beam interaction with the
ultra-strong laser pulse, the initially high value of 
$\chi\approx90$ ensures multiple pair creation. The radiation back-reaction, 
however, splits the initially high value of $(k\cdot p)$ 
between all newborn particles. The reduced values of 
$(k\cdot p)$ result in smaller values of $\chi\propto(k\cdot p)$. 
The cascade terminates, when all particles have
$\chi\le1$ and become incapable of creating new pairs. 

The radiation losses, thereby limit the cascading pair creation. Particularly, emission of softer
$\gamma$ photons 
even may be described within the
radiation force approximation, which is traditionally used to account for the
radiation back-reaction 
(see \cite{ll},\cite{jack},\cite{kogaetal},\cite{ours},\cite{pre}).

The discussed processes are described by the kinetic equations for the
involved particles (electrons,
positrons, $\gamma$-photons). For circularly polarized 1D wave of
constant amplitude, the system of three 1D integro-differential kinetic equations
is reducible to a large system of ODEs, which is solved here numerically.
\section{Electron in QED-strong field} 
The emission probability in the QED-strong 1D wave field may be found in 
Sections
40,90,101 in \cite{lp}, as well as in \cite{nr}. However, to 
simulate highly dynamical effects in pulsed fields, one needs 
a reformulated emission probability, related to short time intervals 
(not $(-\infty,+\infty)$), which is rederived in Appendix A with careful attention to
consistent problem formulation.

Again, the energy, $\hbar\omega^\prime$, and momentum, $\hbar{\bf k}^\prime$,  of the emitted photon are normalized to $m_ec^2$
and $m_ec$. 
The four-dot-product, $(k\cdot p)$, is the 
motional invariant for an electron and it is also conserved in the process of emission:
$(k\cdot p_i)=(k\cdot k^\prime)+(k\cdot p_f)$.  A subscript $i,f$ denotes the 
electron in the initial ({\it i}) or final ({\it f}) state. 

In the 1D wave field the emission probability may be conveniently
related to the interval of the wave phase, $d\xi$, which should be
taken along the electron trajectory. The interval of time, $dt$,
and that of the electron proper time, $d\tau_e$, are
related to $d\xi$ as follows: 
$
d\tau_e=d t/{\cal E}=d\xi/[c(k\cdot p)]$.
The phase volume element for the emitted photon is chosen in the form
$d^2{\bf k}^\prime_\perp d(k\cdot k^\prime)$. The emission probability,
$dW_{fi}/(d\xi d(k\cdot k^\prime))$, is
integrated over $d^2{\bf k}^\prime_\perp$, therefore, it is
related to the element of the phase volume, $d(k\cdot k^\prime)$  
(see detail in Appendix A):
\begin{equation}\label{eq:probabf}
\frac{dW_{fi}}{d(k\cdot k^\prime)d\xi}=
\frac{\alpha\left(\int_{r}^\infty{K_{5/3}(y)dy}+\kappa r
K_{2/3}(r)\right)}{\sqrt{3}\pi\lambdabar_C(k\cdot p_i)^2},
\end{equation}
$$ 
\kappa=\frac{(k\cdot
  k^\prime)\chi_e}{(k\cdot p_i)},\quad r=\frac{(k\cdot
  k^\prime)}{\chi_e(k\cdot p_f)},\quad
\chi_e=\frac32(k\cdot p_i)\left|\frac{d{\bf a}}{d\xi}\right|\lambdabar_C.
$$
Here $K_{\nu}(r)$ is the MacDonald function and $\alpha=e^2/(c\hbar)$. 

{\bf Collision integral.} 
In QED-strong fields we introduce $\chi$-parameter not only for electrons
but also for $\gamma$-photons
and relate the emission probability to
$d\chi_\gamma\propto d(k\cdot k^\prime)$:
\begin{equation}
\chi_{\gamma}=\frac32(k\cdot k^\prime)\left|\frac{d{\bf a}}{d\xi}\right|\lambdabar_C,
\quad
\frac{dW_{fi}}{d\chi_\gamma d\xi}=\alpha
\left|\frac{d{\bf
    a}}{d\xi}\right|U^{e\rightarrow\gamma,e}_{\chi_e\rightarrow\chi_\gamma},
\end{equation}
\begin{equation}
U^{e\rightarrow\gamma,e}_{\chi_e\rightarrow\chi_\gamma}=\frac{\sqrt{3}}{2\pi\chi_e^2}\left[\chi_\gamma
  rK_{2/3}(r)+\int_r^\infty{K_{5/3}(y)dy}\right],
\end{equation}
Here
$r=\chi_\gamma/[\chi_e(\chi_e-\chi_\gamma)]$, $\chi_\gamma\le \chi_e$.  
The electron parameter, $\chi_e$, is taken for the initial state and its
value in the final state is $\chi_e-\chi_\gamma$. 

The distribution functions for electrons and photons may be also integrated over ${\bf
  p}_\perp$ and ${\bf k}^\prime_\perp$ correspondingly. 
Thus integrated functions are distributed over $(k\cdot p)$,
$(k\cdot k^\prime)$. We can parameterize {\it locally} these
distributions via $\chi_e\propto (k\cdot p)$, $\chi_\gamma\propto (k\cdot k^\prime)$
and introduce the 1D distribution functions, $f_e(\chi_e)$ and   
$f_\gamma(\chi_\gamma)$. 

The collision integral (see \cite{pk}) describes the
change in the particle distributions due to emission
and accounts for the electrons, leaving the given phase volume,
$d\chi_e$, and those arriving into it within the
interval,
$d\tilde{\xi}=\alpha|d{\bf a}/d\xi|d\xi=2\alpha c\chi_ed\tau_e/(3\lambdabar_C)$:   
$$
\frac{\delta f_e(\chi_e)}{d\tilde{\xi}}=\int_{\chi_e}^\infty{f_e(\chi)
U^{e\rightarrow\gamma,e}_{\chi\rightarrow\chi-\chi_e}d\chi}-
f_e(\chi_e)\int_0^{\chi_e}{U^{e\rightarrow\gamma,e}_{\chi_e\rightarrow\chi}d\chi},
$$
\begin{equation}\label{eq:collintfull}
\frac{\delta f_\gamma(\chi_\gamma)}{d\tilde{\xi}}=\int_{\chi_\gamma}^\infty{f_e(\chi)
U^{e\rightarrow\gamma,e}_{\chi\rightarrow\chi_\gamma}d\chi}.
\end{equation}

{\bf Radiation force approximation}. One may exclude the emission of softer $\gamma$-photons with
$\chi_\gamma\le\varepsilon\ll1$ from the collision
integral by changing the spans as follows: 
$$
\frac{\delta^{+} f_e(\chi_e)}{d\tilde{\xi}}=\int\limits_{\chi_e+\varepsilon}^\infty{f_e(\chi)
U^{e\rightarrow\gamma,e}_{\chi\rightarrow\chi-\chi_e}d\chi}-
f_e(\chi_e)\int\limits_\varepsilon^{\chi_e}{U^{e\rightarrow\gamma,e}_{\chi_e\rightarrow\chi}d\chi},
$$
\begin{equation}\label{eq:collinttrunc}
\frac{\delta^{+} f_\gamma(\chi_\gamma)}{d\tilde{\xi}}=\int_{\chi_\gamma}^\infty{f_e(\chi)
U^{e\rightarrow\gamma,e}_{\chi\rightarrow\chi_\gamma}d\chi},\qquad\chi_\gamma\ge\varepsilon.
\end{equation}
The excluded process should be treated separately:
\begin{equation}\label{eq:radforce}
\frac{\delta^{(rf)}
f_e(\chi_e)}{d\tilde{\xi}}=\frac{\partial}{\partial \chi_e}
\left[U^{(rf)}_{\chi_e}f_e(\chi_e)\right], 
\end{equation}
\begin{equation}
\frac{\delta^{(rf)}
\int_0^\varepsilon{\chi_\gamma f_\gamma(\chi_\gamma)d\chi_\gamma}}{d\tilde{\xi}}=\int_{0}^\infty{f_e(\chi_e)U^{(rf)}_{\chi_e}d\chi_e},
\end{equation}
where the expression for the {\it radiation force},
\begin{equation}
U^{(rf)}_{\chi_e}=\int\limits_0^\varepsilon{\chi_\gamma U^{e\rightarrow\gamma,e}_{\chi_e\rightarrow\chi_\gamma}d\chi_\gamma},
\end{equation}
is obtained using the standard Fokker-Planck development (see \cite{pk}) of the
collision integral at small $\chi_\gamma\le\varepsilon$:
\begin{eqnarray}
\int\limits_{0}^\varepsilon\left(f_e(\chi_e+\chi_\gamma)U^{e\rightarrow\gamma,e}_{\chi_e+\chi_\gamma\rightarrow\chi_\gamma}-
f_e(\chi_e)U^{e\rightarrow\gamma,e}_{\chi_e\rightarrow\chi_\gamma}\right)d\chi_\gamma\approx\nonumber\\
\approx\frac{\partial}{\partial
  \chi_e}\left(f_e(\chi_e)\int_0^\varepsilon{\chi_\gamma U^{e\rightarrow\gamma,e}_{\chi_e\rightarrow\chi_\gamma}d\chi_\gamma}\right).\nonumber
\end{eqnarray}
The advective term like that in Eq.(\ref{eq:radforce}), once introduced to the kinetic equation,
describes the electron transport along the characteristic lines,
$d\chi_e+d\tilde{\xi}U^{(rf)}_{\chi_e}=0$.
This effect is equivalent to that from an extra four-force term,
$(dp^\mu/d\tau_e)_{\rm rad}$, in the
dynamical equation for the electron four-momentum, $p^\mu$, the
force being such that:
\begin{equation}\label{eq:rfproject}
-U^{(rf)}_{\chi_e}=\frac{d\chi_e}{d\tilde{\xi}}=\frac{\partial \chi_e}{\partial
p^\mu}\left(\frac{dp^\mu}{d\tau_e}\right)_{\rm rad}\frac{d\tau_e}{d\tilde{\xi}}.
\end{equation}
The radiation force is directed along $-p^\mu+k^\mu/(k\cdot p)$. The
two terms describe the energy-momentum lost for radiation and those
absorbed from
the 1D wave field in the course of emission, their total being
orthogonal to $p^\mu$ (see \cite{ours},\cite{pre}).  The force  magnitude may be found
from (\ref{eq:rfproject}): 
$$
\left(
\frac{dp^\mu}{d\tau_e}
\right)_{\rm rad}=-
\frac{2\alpha
c}{3\lambdabar_C}U^{(rf)}_{\chi_e}\left(p^\mu-\frac{k^\mu}{(k\cdot p)}\right).
$$
In the first component of this equation the term $\propto{\cal E}$ controls
the radiation energy loss rate, $I_{\rm QED}$. In dimensional form and related per
time interval, $I_{\rm QED}=-2\alpha m_ec^3
U^{(rf)}_{\chi_e}/(3\lambdabar_C)$. 
At $\chi_e\le\varepsilon\ll1$, $I_{\rm QED}$ tends to the 
expression for the radiation energy loss rate found
within the framework of classical electrodynamics. When the radiation
force approach is generalized for large
$\chi_e\gg1$, the emission spectrum is modified by the QED effects
and only the part of this spectrum (which is minor at $\chi_e\gg1$) is embraced by the radiation force
approximation. 
\section{Photon in QED-strong field} 
The absorption probability for photons in the 1D field is derived in
Appendix B. An electron-positron pair ({\it e,p}) is generated in
the photon absorption  with the conservation law:
$(k\cdot k^\prime)=(k\cdot p_e)+(k\cdot p_p)$.  

The phase volume element for the created electron, again is chosen in the form
$d^2{\bf p}_\perp d(k\cdot p)$. The absorption probability,
$dW_{fi}/(d\xi d(k\cdot p_e))$, is
integrated over the transversal momenta components and
related to the element of the phase volume of electron, $d(k\cdot
p_e)$, resulting in the following collision integral:  
\begin{equation}\label{epminus}
\frac{\delta^- f_{e,p}(\chi_{e,p})}{d\tilde{\xi}}=\int_{\chi_{e,p}}^\infty{f_\gamma(\chi_\gamma)
U^{\gamma\rightarrow e,p}_{\chi_\gamma\rightarrow\chi_e}d\chi_\gamma},
\end{equation}
\begin{equation}\label{gammaminus}
\frac{\delta^- f_\gamma(\chi_\gamma)}{d\tilde{\xi}}=-f_\gamma(\chi_\gamma)\int_0^{\chi_\gamma}{
U^{\gamma\rightarrow e,p}_{\chi_\gamma\rightarrow\chi_e}d\chi_e}.
\end{equation}
Here $r=\chi_\gamma/[\chi_e(\chi_\gamma-\chi_e)]$, $\chi_e=\chi_\gamma-\chi_p\le
\chi_\gamma$ and
\begin{equation}
U^{\gamma\rightarrow e,p}_{\chi_\gamma\rightarrow\chi_e}=\frac{\sqrt{3}}{2\pi\chi_\gamma^2}\left[\chi_\gamma
  rK_{2/3}(r)-\int_r^\infty{K_{5/3}(y)dy}\right].
\end{equation}
\section{Solution for kinetic equations}
As long as the distribution functions are
integrated over the transversal components of momentum and 
expressed in terms of the motional integrals, 
$(k\cdot p_{e,p})$, their evolution is
controlled by the collision integrals:
\begin{equation}\label{eq:tosolve}
\frac{\partial f_{e,p,\gamma}(\tilde{\xi},(k\cdot p_{e,p,\gamma}))}
{\partial\tilde{\xi}}=\left(\delta^{(rf)}+\delta^{+}+
\delta^{-}\right)f_{e,p,\gamma}.
\end{equation}
The derivatives, $\partial/\partial\tilde{\xi}$, are taken  
at constant $(k\cdot p)$. Eqs.(\ref{eq:tosolve}) are easy-to-solve
for the 1D wave field of any shape, however, for 
circularly polarized wave of constant amplitude the solution is especially
simple. In this case $(k\cdot p)$ are different from $\chi$ by a
constant factor, and Eqs.(\ref{eq:tosolve})
may be solved with derivatives,
$\partial/\partial{\tilde{\xi}}$, at constant $\chi$ for the functions, 
$f_{e,p,\gamma}(\tilde{\xi},\chi_{e,p,\gamma})$.

We solve Eqs.(\ref{eq:tosolve}) numerically, by discretizing them 
at a uniform grid, $\chi=I\Delta\chi$,
$I=1,2,3...,N$, with the choice of $\Delta\chi=0.1$, 
$\varepsilon=\Delta\chi/2$. The $\tilde{\xi}$-dependent 
distribution functions at this grid obey the system of $3N$ ODEs, which
is integrated numerically. At initialization, electrons with
$f_e=\delta(\chi_e-\chi_0)$, $\chi_0=90$, 
counterpropagate in the circularly polarized wave field with $|d{\bf
 a}/d\xi|=110$. This choice corresponds to the SLAC electron beam and 
the laser intensity of $J\approx 5\cdot 10^{22}\,{\rm W/cm}^2$ 
for $\lambda=0.8\mu$m,
to be achieved soon. 

\begin{figure}
\includegraphics[scale=0.32]{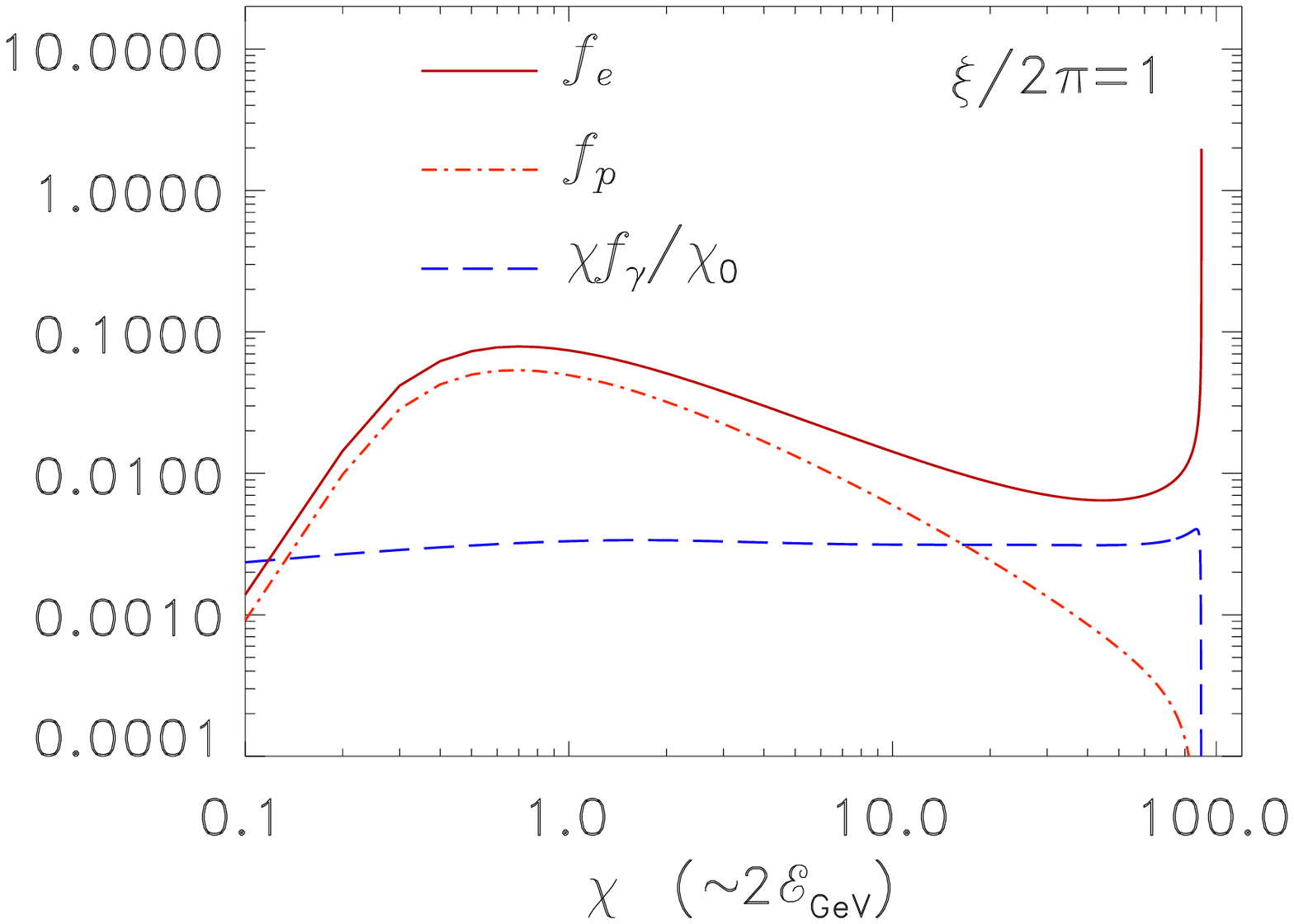} 
\includegraphics[scale=0.32]{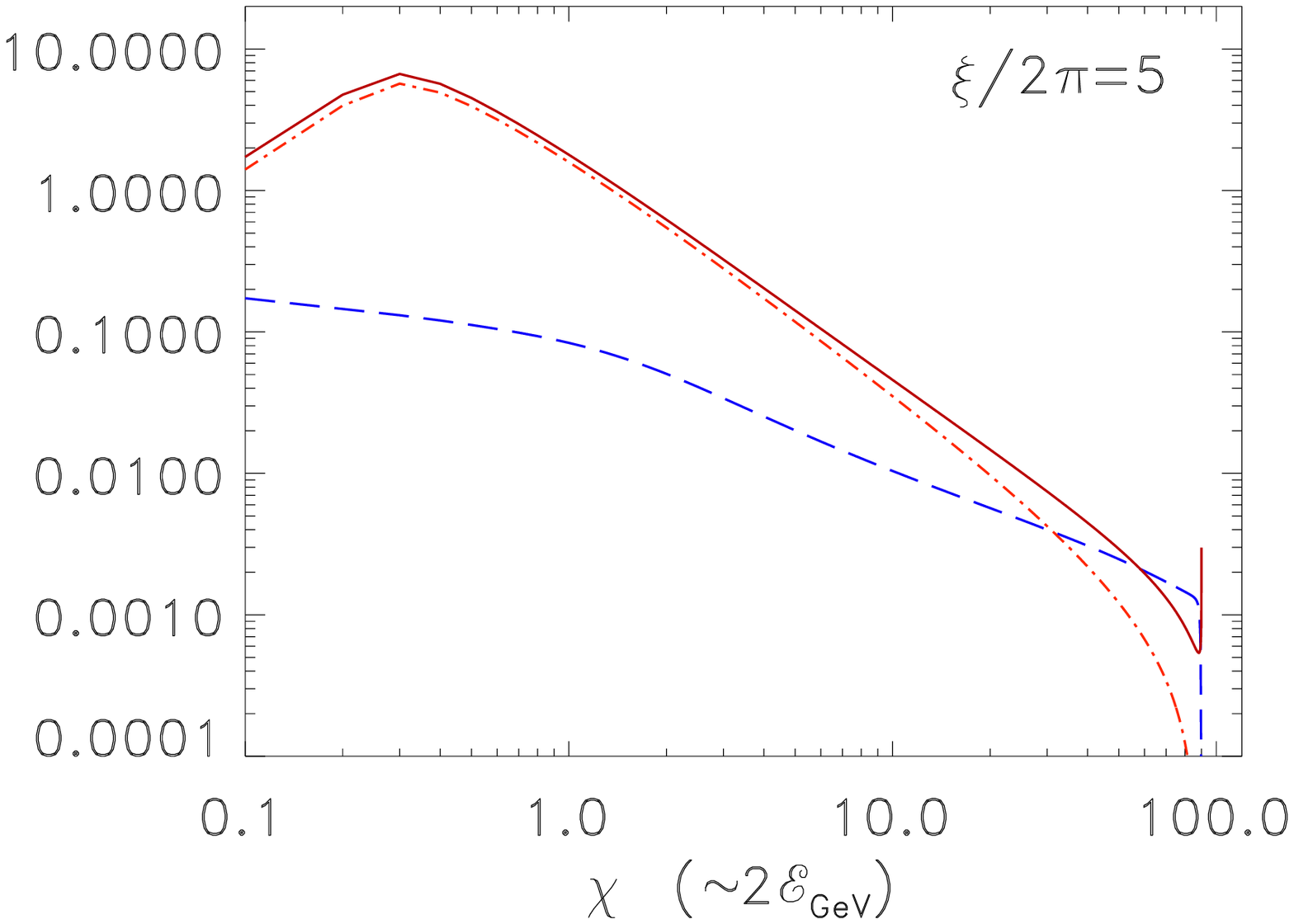} 
\caption{
Distribution functions of electrons and positrons, 
  $f_{e,p}(\chi)$, and a spectrum of emission, $\chi_\gamma
  f_\gamma(\chi)/\chi_0$, after the interaction of 46.6 GeV
  electrons with one cycle (left panel) and five cycles (right panel) of a laser
  pulse of  
intensity  
$J\approx5\cdot 10^{22}\,{\rm W/cm}^2$ (so that
  $\chi\approx2{\cal
  E}[GeV]$ --- see Eq.(\ref{eq:2})). Here $f_e-f_p$ is the 
  distribution of 
the
beam electrons and $\int{(f_e-f_p)d\chi=1}$. 
}
\label{fig1}
\end{figure}   

\begin{figure}
\includegraphics[scale=0.32]{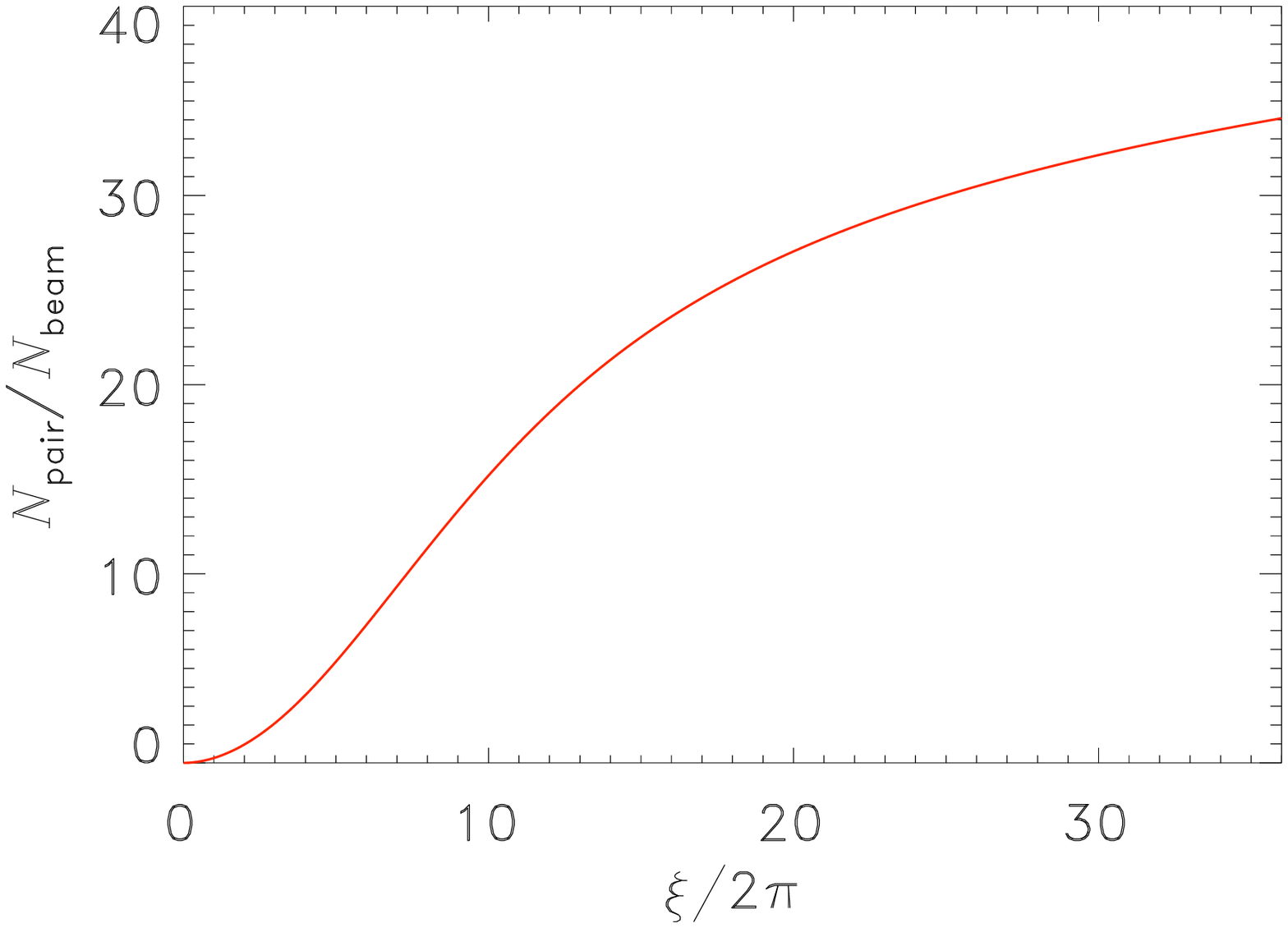} 
\caption{
Pair production for longer pulse 
durations, measured in cycles.
Other
parameters
are the same as 
in Fig.\ref{fig1}.} 
\label{fig2}
\end{figure}

In Fig.\ref{fig1} the beam-wave interaction is traced during 
$\frac\xi{2\pi}=5$ 
cycles of the incident laser pulse 
($\approx13$ fs). The
initial beam electron energy is rapidly converted into $\gamma$-photons
with high $\chi_\gamma$, which then rapidly produce pairs, 
the
typical
rates of the processes being of the order of the inverse light
period. However, the larger fraction of the new particles is born at
$\chi\le 1$, with strongly reduced pair production rate. 
Slow absorption of
photons with $\chi_\gamma\sim1-2$  
maintains
pair production even  
after tens of
wave periods, as shown in Fig.\ref{fig2}.   
\section{Conclusion}
We see that the laser-beam interaction may be accompanied by multiple
pair production. The initial energy of a beam electron is efficiently
spent for creating pairs with significantly lower energies as well as
softer $\gamma$-photons. This effect may be used for producing a
pair plasma. It could also be employed to deactivation
after-use electron beams, reducing radiation hazard. 

The way to solve the kinetic equations is accurate
and it does not employ the Monte-Carlo method. The solution can be used
to benchmark numerical methods designed to simulate
processes in QED-strong laser fields.  

{\bf We acknowledge} help and advice we
received from S.~S.~Bulanov, M.~Hegelich, J.~G.~Kirk, H.~Ruhl,
T.~Schlegel and T.~Tajima.  
One of us (I.S.) is supported by the DOE NNSA under the Predictive
Science Academic Alliances Program by grant DE-FC52-08NA28616.  

\section{Appendix A. Electron in the QED-strong field: emission
  probability}
In weaker fields, especially for 
the particular case of 
a harmonic wave, the emitted power is given by an integral over many
periods of the wave.  
In this case,
the solution of the emission
problem in the weak harmonic wave field (under the requirement on the
wave amplitude opposite to that formulated in Ineq.(\ref{eq:strong})) is 
given in 
Section 101 of 
\cite{lp} as a sum over multi-photon-orders,
resulting from the Fourier-series expansion for the (infinitely long)
periodic wave. This standard approach, however, may become
meaningless as applied 
to 
ultra-strong laser
pulses, for many reasons. These pulses may be so short that they cannot 
be thought of as 
harmonic waves. 
Their fields may be strong enough to force an
electron to expend its energy on radiation  
in less
than a single
wave period. However, an even more important point is that the radiation
loss rate and even the spectrum of radiation is no longer an integral 
characteristic of the particle motion through a number of wave periods:
a local dependence of emission on both particle and field
characteristics is typical for the strong fields.  The latter statement may
be found in \cite{lp},  
Section 101.
For the particular case of the 1D wave
field the evaluation of the {\it formation time} for 
emission 
is
provided in \cite{pre} within the framework of classical 
electrodynamics. 
It  
is
shown that the formation time is much shorter
than the wave period as long as Ineq.(\ref{eq:strong}) is fulfilled.   

Here the emission problem is discused 
for 
QED-strong fields.
We consider  
a 1D wave field  
taken in the Lorenz gauge \cite{okun}: 
$$
a^\mu=a^\mu(\xi),\qquad \xi=(k\cdot x),\qquad(k\cdot a)=0,
$$  
$a^\mu=(0,{\bf a})$, $k^\mu$ and $x^\mu$ being the
4-vectors of the 
potential, the wave and the coordinates. Herewith
the 4-dot-product is introduced in a 
usual manner: 
$(k\cdot x)=k^\mu x_\mu = \omega t-({\bf k}\cdot{\bf x})$  
etc. Space-like 3-vectors (i.e., the first to the third components of
a 4-vector)  in contrast with 4-vectors are denoted in bold,
4-indices are denoted with Greek letters. Recall, that a 
metric signature $(+,-,-,-)$ is used, therefore, for
space-like vectors the 3D scalar product and 4-dot-product have
opposite signs, particularly:
$$
\left(\frac{d{\bf a}}{d\xi}\right)^2=
-\left(\frac{da}{d\xi}\right)^2\ge 0.
$$    

\subsection{Transformed space-time}
A 
method facilitating 
many derivations 
involves the introduction of a specific time-space coordinate frame.   
Introduce a 
Transformed Space-Time (TST) :  
$$
x^{0,1}=(ct\mp x_\|)/\sqrt{2},\qquad x^{2,3}={\bf x}_\perp,
$$ 
subscript $\perp$ denoting the vector components 
orthogonal to ${\bf k}$. The properties of the TST provide a convenient
description for the classical motion of an electron in the 1D wave
field. Note first that
$$
dx^0=\frac{\lambdabar d\xi}{\sqrt{2}}, \qquad
p^0=\frac{\lambdabar(k\cdot p)}{\sqrt{2}}, \qquad (p\cdot k)= 
\frac{{\cal E}-p_\|}{\lambdabar}.
$$ 
Second,  
the generalized momentum components,  
$p^0$ and ${\bf p}_{\perp 0}={\bf p}_\perp+{\bf a}$,  
are conserved. Third, the metric tensor in the TST is:
$$
G^{01}=G^{10}=1,\qquad G^{22}=G^{33}=-1, \qquad G^{\mu\nu}=G_{\mu\nu}. 
$$
Note 
the
unusual off-diagonal structure of the metric tensor, resulting
in a strange relationship between 
contravariant and covariant
coordinates: $x^0=x_1$, $x^1=x_0$. Specifically, the (contravariant)
component of the electron momentum, $p^0$, is a motional invariant, 
as long as 
the
vector-potential (and the Hamiltonian, $H$) does not depend on the (covariant)
coordinate, $x_0$ (despite 
its dependence 
on $x^0$): 
$$
\frac{dp^0}{dt}\propto-\frac{\partial H}{\partial x_0}=0.
$$
Finally, the identity, ${\cal E}^2=p^2+1$, being expanded in the TST metric, gives: 
$$
p^1=\frac{1+{\bf p}_\perp^2}{2p^0}=
\frac{1+({\bf p}_{\perp 0}-{\bf a})^2}{\sqrt{2}\lambdabar(k\cdot p)}.
$$  
The derivative over $x^0$ or, the same, over $\xi$ is conveniently
related to the derivative over the proper time for electron: 
\begin{equation}
\frac{d}{d\tau}={\cal E}\left[\frac\partial{\partial t}+({\bf
    v}\cdot\frac\partial{\partial {\bf x}})\right]=c(k\cdot
p)\frac{d}{d\xi}.
\end{equation}
\subsection{Classical trajectory and momenta retarded product}
Many characteristics of emission may be expressed in terms of 
the 
relationship between 
the
4-momenta of 
the
electron
at different 
instants:
\begin{equation}\label{eq:classic}
p^\mu(\xi)=p^\mu(\xi_1)-\delta a^\mu+
\frac{2(p(\xi_1)\cdot \delta a)-(\delta a)^2}{2(k\cdot p)}k^\mu,
\end{equation}
where 
$$\delta a^\mu=a^\mu(\xi)-a^\mu(\xi_1).$$
As a 
consequence 
from Eq.(\ref{eq:classic}), one can obtain the expression for the {\it Momenta Retarded Product} (MRP):
\begin{equation}\label{eq:pp}
(p(\xi)\cdot p(\xi_1))=1-\frac{(\delta a)^2}2=1+\frac{(\delta{\bf a})^2}2,
\end{equation}

Note, that the MRP is given by Eq.(\ref{eq:pp}) for an  
arbitrary difference 
between $\xi$ and $\xi_1$, but only for 
the particular case of the 1D
wave field. However the limit of this formula 
as 
$|\xi-\xi_1|\rightarrow 0$, which is as follows:
$$
(p(\xi)\cdot p(\xi_1))|_{|\xi-\xi_1|\rightarrow 0}\approx 
1+ \frac12(\xi-\xi_1)^2\left|\frac{d{\bf a}}{d\xi}\right|^2
$$
or, in terms of the 
MRP in the proper time, $\tau$:
\begin{equation}\label{eq:gencovariance}
(p(\tau)\cdot p(\tau+\delta\tau))=1-(\delta\tau)^2\frac{(f\cdot f)}{2m_e^2c^2},
\end{equation}
has a much wider range of applicability. Eq.(\ref{eq:gencovariance}) is
derived from the equation of
motion:
$$
\frac{dp^\mu}{d\tau}=\frac{f^\mu}{m_ec},
$$
using the identities: 
$$
(p(\tau)\cdot p(\tau))=1,\qquad (p(\tau)\cdot f(\tau))=0,
$$
$$
\frac{d(p(\tau)\cdot p(\tau+\delta\tau))}{d(\delta\tau)}=-\delta\tau\left(\frac{dp}{d\tau}\cdot\frac{dp}{d\tau}\right)+O((\delta\tau)^2).
$$
Here $f^\mu$ is the Lorentz four-force:
$$
f^\mu = (f^0,{\bf f}^{(3)}),\qquad f^0=e{\bf E}\cdot{\bf p},\qquad 
{\bf f}^{(3)}=e{\cal E}{\bf E}+e{\bf p}\times{\bf B}.
$$
\subsection{A solution of the Dirac equation} 
The Dirac equation which determines the evolution of the wave function, 
$\psi$, for a {\it non-emitting} electron in the external field, reads:
\begin{equation}\label{eq:Dirac}
\left[i \lambdabar_C\left(\gamma\cdot\frac\partial{\partial
  x}\right)-(\gamma\cdot a)\right]\psi=\psi,
\end{equation}
$\gamma^\mu$ being the Dirac $4*4$ matrices,
$(\gamma^0,\gamma^1,\gamma^2,\gamma^3)$. The relativistic dot-product
of the Dirac matrices by 4-vector, such as $(\gamma\cdot a)$, is the
linear combination of the Dirac matrices: 
$(\gamma\cdot a)=\gamma^0a^0-\gamma^1a^1-\gamma^2a^2-\gamma^3a^3$. Such
a linear combination, which is also
a
$4*4$ matrix, may be multiplied
by another matrix of this kind or by 4-component bi-spinor, such as $\psi$,
following 
matrix multiplication rules. For example, $(\gamma\cdot
a)\psi$ is a bi-spinor, as
is
the matrix, $(\gamma\cdot a)$ multiplied from the right
hand side by the bi-spinor, $\psi$. 

The solution of Eq.(\ref{eq:Dirac}) in a form of a plane
electron wave can be conveniently expressed in terms of the classical solution:
\begin{equation}
\psi=\frac1{\sqrt{N}} u(p(\xi)) \exp(-i\int^x{\left((p(\xi)+eA)\cdot dx\right)}),
\end{equation}
the normalization coefficient $N={\rm const}$. 
By expanding the phase multiplier in the TST, a more convenient form can be provided: 
\begin{equation}\label{eq:volkov}
\psi=\frac{ u(p(\xi))P(\xi)}{\sqrt{N}}
\exp\left[
\frac{i
\left[
({\bf p}_{\perp0}\cdot {\bf x}_\perp)-
\frac{\lambdabar(k\cdot p)x^1}{\sqrt{2}}
\right]
}{\lambdabar_C}
\right].
\end{equation}

Here
$u(p(\xi))$ is plane wave bi-spinor amplitude, which satisfies the
system of four linear algebraic equations: 
\begin{equation}\label{bispinor}
(\gamma\cdot p(\xi))u(p(\xi))=u(p(\xi)),
\end{equation}
as well as the normalization condition:
 $\hat{u}u=2.$ 
The  $\xi$-dependent phase multiplier, $P(\xi)$, is as follows:
$$
P(\xi)=\exp\left(-\frac{i}{\lambdabar_C}
\int^{\xi}{\frac{1+{\bf p}_\perp^2(\xi_2)}{2(k\cdot p)}
d\xi_2}\right),
$$
or,
\begin{equation}\label{xidependent}
P(\xi)=P(\xi_1)\exp\left(-\frac{i}{\lambdabar_C}
\int^{\xi}_{\xi_1}{\frac{1+{\bf p}_\perp^2(\xi_2)}{2(k\cdot p)}
d\xi_2}\right).
\end{equation} 
Using Eq.(\ref{eq:classic}), one can find:
\begin{equation}\label{eq:QEDpropagator}
u(p(\xi))=\left[1+\frac{(\gamma\cdot
 k)\left(\gamma\cdot [a(\xi)-a(\xi_1)]
\right)}{2(k\cdot p)}\right]u(p(\xi_1))
\end{equation}  
and verify that Eq.(\ref{eq:volkov}) satisfies the Dirac equation. 
To prove all these assertions, note that the plane wave bi-spinor
amplitude once expressed in terms
of the classical solution Eq.(\ref{eq:classic}), satisfies the following 
commutation rule:
\begin{equation}\label{eq:commutation}
\left[(\gamma\cdot p(\xi))\pm 1\right]\left[1+\frac{(\gamma\cdot k)}
{2(k\cdot p)}(\gamma\cdot \delta a)\right]=
\left[1+\frac{(\gamma\cdot k)}
{2(k\cdot p)}(\gamma\cdot \delta a)\right]\left[(\gamma\cdot p(\xi_1))\pm 1\right].
\end{equation}
The latter can be proved using the commutation rules as in Eqs.(22.5)
from \cite{lp} for the Dirac matrices 
as well as the identities, $(k\cdot A)=(k\cdot k)=0$: 
$$
\left[(\gamma\cdot p(\xi_1))
-(\gamma\cdot \delta a)+
\frac{2(p(\xi_1)\cdot \delta a)-(\delta a)^2}{2(k\cdot p)}(\gamma\cdot k) \pm 1\right]
\left[1+\frac{(\gamma\cdot k)(\gamma\cdot \delta a)}{2(k\cdot p)}
\right]= 
$$
$$
(\gamma\cdot p(\xi_1))
-(\gamma\cdot \delta a)+
\frac
{2(p(\xi_1)\cdot \delta a)-
(\gamma\cdot p(\xi_1))(\gamma\cdot \delta a)}{2(k\cdot p)}(\gamma\cdot k)
\pm \left[1+\frac{(\gamma\cdot k)(\gamma\cdot \delta a)}{2(k\cdot p)}
\right]=
$$
$$
=\left[1+\frac {(\gamma\cdot k)(\gamma\cdot \delta a)}{2(k\cdot p)}\right]
\left[(\gamma\cdot p(\xi_1)) \pm 1\right].
$$
Thus, Eq.(\ref{eq:commutation}) is proved. This commutation rule allows us to
verify that $u(p(\xi))$ is indeed a bi-spinor amplitude of a planar
wave with the four-vector $p(\xi)$, satisfying Eq.(\ref{bispinor}). 
Finally, to verify that Eq.(\ref{eq:volkov}) gives the solution of the
Dirac equation, one should account for Eq.(\ref{bispinor}) as
well as the identity, 
$$(\gamma\cdot k)\frac{d[u(p(\xi))]}{d\xi}=(\gamma\cdot k)(\gamma\cdot
k)\left(\gamma\cdot \frac{dA}{d\xi}\right) u(p_0)=0,$$ 
which is valid, 
because $(\gamma\cdot k)(\gamma\cdot k)=k^2=0$, 

The 
advantage of the 
approach used here  
as compared to the known
Volkov solution (see 
Section 40
in \cite{lp}) is that the wave
function in Eqs.(\ref{eq:volkov}-\ref{eq:QEDpropagator})
is described in a self-contained manner within some finite time
interval, $(\xi_1,\xi)$ (in fact, 
this interval is assumed 
to be very short below) in terms of the local parameters of the classical 
trajectory of electrons. This 
approach is better applicable to
strong fields, in which the time interval between subsequent emission
occurrences, which destroys the unperturbed wave function, becomes very short.
\subsection{The matrix element for emission} 
The emission problem is formulated in the following way. The electron
motion in the strong field may be thought of as a sequence of short
intervals. Within each of these intervals the electron follows a piece
of a classical trajectory, as in Eq.(\ref{eq:classic}), and its wave
function (an electron state) is given by Eq.(\ref{eq:volkov}). The 
transition from one piece of the classical trajectory to another, or, 
the same, from one electron state
to another occurs in a probabilistic manner. The probability of this
transition, which is accompanied by a photon emission is calculated
below using the QED perturbation theory. 
We consider the laser wave
field as a classical field ($\lambdabar_C(k\cdot p)\ll 1$) and 
apply
first-order
 perturbation theory with respect to the 
Hamiltonian
of the electron-photon interaction, 
$\hat{H}\propto(\hat{j}\cdot\hat{A})$.

The only difficulty specific
to strong pulsed fields is that the short piece of the electron
trajectory is strictly bounded in space and in time, while the QED
invariant perturbation theory is based on the 'matrix element', which
is the integral over infinite 4-volume.       
\begin{figure}
\includegraphics[scale=0.4, angle=90]{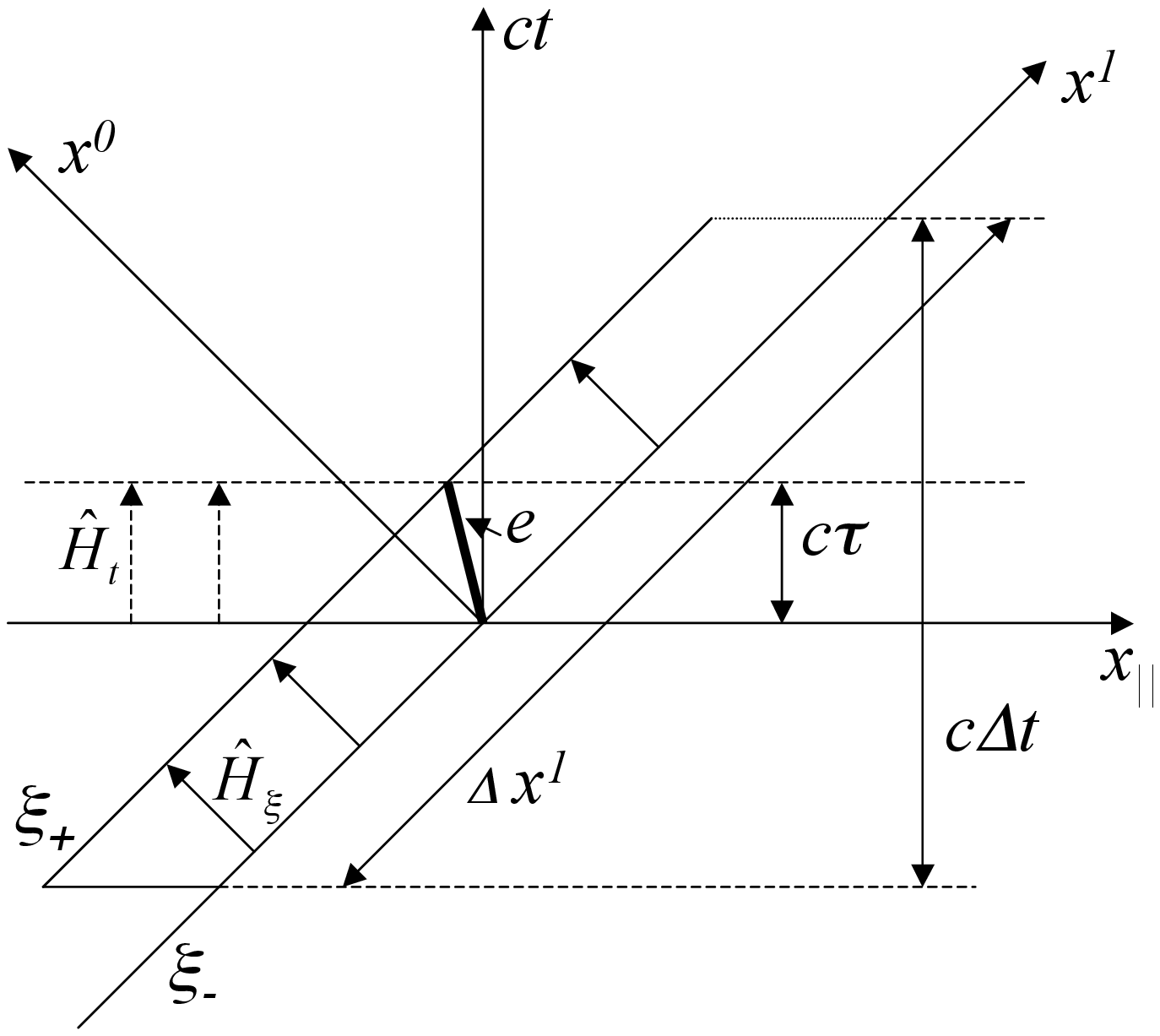} 
\caption{The volume over which to integrate the matrix element while finding 
the emission probability: in the standard scheme for the dipole emission 
(in dashed lines) and in the TST (in solid lines). Arrows show the 
direction, 
along which the Heisenberg operator advances the wave functions.
}
\label{fig_1}
\end{figure} 

To avoid this difficulty the following method is suggested, which is
analogous to the dipole emission theory as applied in TST. Introduce
domain, $\Delta^4x=(\Delta x^1*S_\perp)*\Delta x^0$, 
bounded by two hypersurfaces, $\xi=\xi_{-}$ and $\xi=\xi_+$
(see Fig.\ref{fig_1}). The difference $\xi_+-\xi_-$ 
is bounded as described below,
so that $\Delta^4x$ covers only a minor part of the pulse.   
A volume, 
$$
V=S_\perp\lambdabar(\xi_+-\xi_-)
=S_\perp\lambdabar\int_{\xi_-}^{\xi_+}{d\xi_2},$$ 
is a section of $\Delta^4x$
subtended by a line $t={\rm const}$. 

With the following choice for the normalization coefficient in Eq.(\ref{eq:volkov}): 
$$
N=2S_\perp \lambdabar\int_{\xi_-}^{\xi_+}{{\cal E}
(\xi_2)d\xi_2},
$$  
the integral of the electron density in the volume $V$ 
$$
\int{\hat{\psi}\gamma^0\psi dV}=
S_\perp \lambdabar\int_{\xi_-}^{\xi_+}{\hat{\psi}\gamma^0\psi
  d\xi_2},$$ 
is set to unity, i.e. there is a single electron in the volume
$V$. This statement follows from  Eq.(\ref{eq:volkov}) and the known
property of normalized bi-spinor amplitudes:
$\hat{u}\cdot\gamma^0 \cdot u=2{\cal E}$. Here the hat means the Dirac conjugation.
 
For a photon of 
wave vector, $(k^\prime)^\mu$, and polarization vector, $l^\mu$,
introduce the 
wave function:
$$
(A^\prime)^\mu=
\frac{
\exp[-i(k^\prime\cdot x)/\lambdabar_C]
     }
{
 \sqrt{N_p}
}l^\mu,
$$
or, by expanding this in the TST:
$$
(A^\prime)^\mu=
\frac{P_p(\xi)}{\sqrt{N_p}}
\exp\left[
\frac{i({\bf k}^\prime_\perp\cdot {\bf x}_\perp)}{\lambdabar_C}-
\frac{
i\lambdabar (k\cdot k^\prime) x^1}
{\sqrt{2}\lambdabar_C}
     \right]
l^\mu,$$
where:
$$P_p(\xi)=
\exp\left[-i \xi
\frac{ ({\bf k}^\prime_\perp)^2}
{2(k\cdot k^\prime)\lambdabar_C}
\right].
$$
Here the photon momentum and photon energy are related to $m_ec$ and
$m_ec^2$ correspondingly, or, equivalently, dimensionless
$(k^\prime)^\mu$ equals dimensional $(k^\prime)^\mu$ multiplied by
$\lambdabar_C$. 
The choice of the normalization coefficient, 
$$
N_p=\frac{\omega^\prime V}{2\pi\hbar c\lambdabar_C},
$$ 
corresponds to a
single photon in the volume, $V$.

The emission probability,  $dW$,  
is given by an integral  
over $\Delta^4x$:
\begin{equation}\label{eq:probab}
dW=\frac{\alpha L_fL_p}{\hbar c}
\left|\int{\hat{\psi}_f(\gamma\cdot (A^\prime)^*)\psi_idx^0dx^1dx^2dx^3}\right|^2.
\end{equation} 
Here
\begin{equation}\label{eq:firstphase}
L_p=\frac{Vd^3{\bf k}^\prime}{(2\pi\lambdabar_C)^3}=
\frac{\hbar c N_p d^2{\bf k}^\prime_\perp d(k\cdot
k^\prime)}{(2\pi\lambdabar_C)^2(k\cdot k^\prime)}
\end{equation} 
is the number of states for the emitted photon. The transformation of
the phase volume as in Eq.(\ref{eq:firstphase}) is based on the
following Jacobian:
$$
\left(\frac{\partial k^\prime_\|}{\partial(k^\prime\cdot
  k)}\right)_{{\bf k}^\prime_\perp={\rm const}}=\frac{\omega^\prime}{(k^\prime\cdot
  k)},
$$
which is also used below in
many places. A subscript $i,f$ denotes the 
electron in the initial ({\it i}) or final ({\it f}) state.  
The number of electron states in the presence of the wave field, 
$L_{i,f}$, should be integrated over the volume $V$  
$$
L_{i,f}=\frac{1}{(2\pi\lambdabar_C)^3}\int_V{d^3{\bf p}_{i,f}dV}=
\frac{d(k\cdot p)_{i,f}d^2{\bf p}_{\perp i,f}N_{i,f}}
{2(2\pi)^3\lambdabar_C^3(k\cdot p)_{i,f}}.
$$
\subsection{Conservation laws} 
The integration by $dx^1dx^2dx^3=c\sqrt{2} dtd^2{\bf x}_\perp$  
results in three $\delta-$ functions, expressing the conservation of 
totals of ${\bf p}_\perp$ and $(k\cdot
p)$, for particles in initial and 
final states:
$$
{\bf
  p}_{\perp i}={\bf
  p}_{\perp f}+{\bf
  k}_\perp^\prime,\qquad (k\cdot p_i)=(k\cdot p_f)+
  (k\cdot k^\prime).
$$
Twice integrated with respect to $dx^1$, the probability $dW$ is  
proportional to a long time interval, $\Delta t=\Delta x^1/(c\sqrt{2})$,  
if the boundary condition for the electron wave at $\xi=\xi_-$
is maintained within that long time. On transforming the integral over
$dx^0$ to that over $d\xi$, one can find:
$$
\left|\int{...d^4x}\right|^2 =(2\pi\lambdabar_C)^3 S_\perp c\Delta t\lambdabar\left|\int{...d\xi}\right|^2\times
$$
$$
\times \delta^2({\bf
  p}_{\perp i}-{\bf
  p}_{\perp f}-{\bf
  k}_\perp^\prime)\delta((k\cdot p_i)-(k\cdot p_f)-
  (k\cdot k^\prime)).
$$
To take the large value of $\Delta t$ seems to be the only way to
calculate the integral, however, the emission probability calculated in
this way relates to multiple electrons in the initial state, each of
them locating  between the wave fronts
$\xi=\xi_-$ and $\xi=\xi_+$ during much shorter time, 
\begin{equation}\label{eq:time}
\delta t(\xi_-,\xi_+)=(1/c)\int_{\xi_-}^{\xi^+}{{\cal E}_i(\xi)d\xi_2}/(k\cdot p_i). 
\end{equation}
For a single electron the emission probability becomes: 
$$dW_{fi}(\xi_-,\xi_+)=\delta t dW/\Delta t.$$  
Using $\delta-$ functions
it is easy to integrate Eq.(\ref{eq:probab}) over $d{\bf p}_{\perp f}d(k\cdot
p_f)$:
$$
\frac{dW_{fi}(\xi_-,\xi_+)}{d(k\cdot k^\prime)d^2{\bf k}^\prime_\perp}=
\frac{\alpha \left|
   \int_{\xi_-}^{\xi_+}{T(\xi)\hat{u}(p_f)(\gamma\cdot l^*)u(p_i)d\xi
       }\right|^2}{(4\pi\lambdabar_C)^2(k\cdot k^\prime)(k\cdot
  p_i)(k\cdot p_f)},
$$
where 
$$
T(\xi)=\frac{P_i(\xi)}{P_f(\xi)P_p(\xi)}=
\exp\left[\frac{i\int^\xi{(k^\prime\cdot p_{i,f}(\xi_2))d\xi_2}}{\lambdabar_C(k\cdot p_{f,i})}
\right],
$$ 
$P_i(\xi)$ and $P_f(\xi)$ are the electron phase multipliers, $P(\xi)$,
for the electron in initial and final states and 
\begin{equation}\label{eq:cons}
p^\mu_f(\xi)=p^\mu_i(\xi)-(k^\prime)^\mu+\frac{(k^\prime\cdot p_i(\xi))}
{(k\cdot p_i)-(k\cdot k^\prime)}k^\mu . 
\end{equation}
If in the expression for $T(\xi)$ the numerator of the fraction is
taken for the initial state of electron, then the denominator should be
taken for the final state and vise versa.

Prior to discussing Eq.(\ref{eq:cons}), return to
Eq.(\ref{eq:classic}) and analyze it component-by-component in the
TST. It appears that three of the four components of that equation
describe the conservation of 
$(k\cdot p)$ and ${\bf p}_{\perp 0}={\bf p}_{\perp}+{\bf a}$ 
for electron {\it in the course of its
emission-free motion}. At the same time, yet another component of 
Eq.(\ref{eq:classic}), specifically, $p^1$, directed along $k^\mu$, describes the
energy-momentum exchange between the electron and the 1D wave field, maintaining the
identity, $(p\cdot p)=1$. Now turn to Eq.(\ref{eq:cons}). Again, three
of the four components express the conservation of the same variables
{\it in the course of the photon emission}, while
the
 $p^1$ component,
directed along $k^\mu$ describes the absorption of energy and momentum
from the wave field in the course of the photon emission. Note, that in
the case of a strong field, the energy absorbed from field is not an
integer number of quanta, and that for short non-harmonic field it is
not even  
a constant, but a function of the local field.
\subsection{Calculation of the matrix element: a product of the Dirac matrices}
To 
calculate the matrix element, one can re-write it as the 
double integral over $d\xi d\xi_1$. The integrand in this
double integral includes the phase multiplier and the following
matrix product: 
\begin{equation}\label{eq:matproduct}
\hat{u}(p_f(\xi))(\gamma\cdot l^*)u(p_i(\xi))
       \left[\hat{u}(p_f(\xi_1))(\gamma\cdot
         l^*)u(p_i(\xi_1))\right]^*=...
\end{equation} 
Using Eq.(\ref{eq:QEDpropagator}) one can reduce the multipliers of the kind of
$u(p_i(\xi))\otimes \hat{u}(p_i(\xi_1))$ to $u(p_i(\xi_1))\otimes
\hat{u}(p_i(\xi_1))$,
which can be expressed in terms of the polarization density matrix at the position 
$\xi_1$. Although in a strong wave 
electrons may be polarized (see \cite{Omori}),  
in the present work the emission probability is assumed to be averaged
over the electron initial and
final polarizations. Therefore, the polarization matrix is used in the
form of $1/2[(\gamma\cdot p_i(\xi))+1]$ and one can re-write the integrand
as follows:
$$
...=\frac14{\rm Sp}\{
\left[1+\frac{(\gamma\cdot k)}{2(k\cdot p_f)}
(\gamma\cdot[a(\xi_1)-a(\xi)])\right]\times
$$
$$
\times
\left[(\gamma\cdot p_f(\xi))+1\right](\gamma\cdot l^*)\times\nonumber\\
\left[1+\frac{(\gamma\cdot k)}{2(k\cdot p_i)}
(\gamma\cdot [a(\xi)-a(\xi_1)])\right]
\left[(\gamma\cdot p_i(\xi_1))+1\right](\gamma\cdot l)    
\}=...
$$  
In transforming the trace of matrix in addition to the mentioned above 
relationships and the Dirac matrix algebra we use the commutation rule 
Eq.(\ref{eq:commutation}), the conservation law Eq.(\ref{eq:cons}), and the identity $(l^*\cdot l)=-1$ for a spatial
unity vector of the emitted photon polarization, which can be also written as
$(\gamma \cdot l^*)(\gamma\cdot p_{i,f})(\gamma\cdot l)=2(l^*\cdot p_{i,f})
(\gamma\cdot l)+(\gamma\cdot p_{i,f})$. 
Move $(\gamma\cdot p_f(\xi)+1)$ to the first position using the commutation 
rule Eq.(\ref{eq:commutation}) and then move it to the last position. In the product 
$[(\gamma\cdot p_i(\xi_1))+1](\gamma\cdot l)[(\gamma\cdot p_f(\xi_1))+1]$ 
we should keep only odd powers in $\gamma$, as long as in the
preceding multiplier only odd powers of $\gamma$ are present:
$$
...=\frac14 {\rm Sp}\{
\left[1+\frac{(\gamma\cdot k)}{2(k\cdot p_f)}
(\gamma\cdot[a(\xi_1)-a(\xi)])\right]
(\gamma\cdot l^*)\times
$$
$$
\left[1+\frac{(\gamma\cdot k)}{2(k\cdot p_i)}
(\gamma\cdot[a(\xi)-a(\xi_1)])\right]
\left[(\gamma\cdot p_i(\xi_1))(\gamma\cdot l)(\gamma\cdot p_f(\xi_1))+(\gamma\cdot l)\right]     
\}=
$$
In the product of the first three multipliers separate the even powers in $a$, 
which are 
$(\gamma\cdot l^*)-(\gamma\cdot k)(l^*\cdot
k)(a(\xi)-a(\xi_1))^2/[2(k\cdot p_i)(k\cdot p_f)]$. The traces of the
products of two or four Dirac matrices are calculated using
Eqs.(22.9-10) from \cite{lp}. 
$$
=\left[(p_i(\xi_1)\cdot p_f(\xi_1))-1\right]
\left[1+\frac{(l^*\cdot k)(l\cdot k)}{2(k\cdot p_i)(k\cdot p_f)}
(a(\xi)-a(\xi_1))^2\right]+(l^*\cdot p_i(\xi_1))
(l\cdot p_f(\xi_1))+
$$
$$
+(l\cdot p_i(\xi_1))
(l^*\cdot p_f(\xi_1))
-\left[\frac{(l^*\cdot k)(l\cdot p_i(\xi_1))}{2(k\cdot p_i)}+\frac{(l^*\cdot k)(l\cdot p_f(\xi_1))}{2(k\cdot p_f)}\right](a(\xi)-a(\xi_1))^2+
$$
In one of the two terms linear in $a$ we move 
$(\gamma\cdot l^*)$ from the first to the last position:
$$
+\frac1{8(k\cdot p_i)}{\rm Sp}\left\{(\gamma\cdot k)(\gamma\cdot [a(\xi)-a(\xi_1)])(\gamma\cdot p_i(\xi_1))[2(l^*\cdot p_f(\xi_1))(\gamma \cdot l)+
(\gamma\cdot p_f(\xi_1))]\right\}-
$$
$$
- 
\frac1{8(k\cdot p_f)}{\rm Sp}\left\{(\gamma\cdot k)(\gamma\cdot [a(\xi)-a(\xi_1)])
[2(l^*\cdot p_i(\xi_1))(\gamma \cdot l)+(\gamma\cdot p_i(\xi_1))]
(\gamma\cdot p_f(\xi_1)) \right\}=
$$
Calculate traces for the products of four Dirac matrices in the terms
linear in $a$:
$$
=\left[(p_i(\xi_1)\cdot p_f(\xi_1))-1\right]
\left[1+\frac{(l^*\cdot k)(l\cdot k)}{2(k\cdot p_i)(k\cdot p_f)}
(a(\xi)-a(\xi_1))^2\right]+(l^*\cdot p_i(\xi_1))
(l\cdot p_f(\xi_1))+
$$
$$
+ (l\cdot p_i(\xi_1))
(l^*\cdot p_f(\xi_1))-\left[\frac{(l^*\cdot k)(l\cdot p_i(\xi_1))}{2(k\cdot p_i)}+
\frac{(l^*\cdot k)(l\cdot p_f(\xi_1))}{2(k\cdot p_f)}\right]
(a(\xi)-a(\xi_1))^2+
$$
$$
+\{-\frac 12(\left[a(\xi)-a(\xi_1)\right]\cdot p_i(\xi_1))+
$$
$$
+\frac{(k\cdot p_i)}{2(k\cdot p_f)}
(\left[a(\xi)-a(\xi_1)\right]\cdot p_f(\xi_1))
-\frac 12(\left[a(\xi)-a(\xi_1)\right]\cdot p_f(\xi_1))+
\frac{(k\cdot p_f)}{2(k\cdot p_i)}
(\left[a(\xi)-a(\xi_1)\right]\cdot p_i(\xi_1))-
$$
$$
-\left(\left[p_f(\xi_1)+p_i(\xi_1)\right]\cdot l^*\right)
\left(l\cdot\left[a(\xi)-a(\xi_1)\right]\right)+
\frac{([a(\xi)-a(\xi_1)]\cdot p_i(\xi_1))(k\cdot l)(l^*\cdot p_f(\xi_1))}{(k\cdot p_i)}+
$$
$$
\frac{([a(\xi)-a(\xi_1)]\cdot p_f(\xi_1))(k\cdot l)(l^*\cdot p_i(\xi_1))}{(k\cdot p_f)}\}=
$$
Group the terms proportional to $(l^*\cdot p_{i,f}(\xi_1))$
$$
=(l^*\cdot p_{i}(\xi_1))\left(l\cdot\left\{p_f(\xi_1)-\left[a(\xi)-a(\xi_1)\right]+k\frac{([a(\xi)-a(\xi_1)]\cdot p_f(\xi_1))}{(k\cdot p_f)}\right\}\right)+
$$
$$
+(l^*\cdot p_{f}(\xi_1))\left(l\cdot\left\{p_i(\xi_1)-\left[a(\xi)-a(\xi_1)\right]+k\frac{([a(\xi)-a(\xi_1)]\cdot p_i(\xi_1))}{(k\cdot p_i)}\right\}\right)+
$$
Group in an analogous manner the terms, proportional to 
$(p_{i,f}(\xi_1)\cdot...)$:
$$
+\frac12\left(p_{i}(\xi_1)\cdot\left\{p_f(\xi_1)-\left[a(\xi)-a(\xi_1)\right]+k\frac{([a(\xi)-a(\xi_1)]\cdot p_f(\xi_1))}{(k\cdot p_f)}\right\}\right)+
$$
$$
+\frac12\left(p_{f}(\xi_1)\cdot\left\{p_i(\xi_1)-\left[a(\xi)-a(\xi_1)\right]+k\frac{([a(\xi)-a(\xi_1)]\cdot p_i(\xi_1))}{(k\cdot p_i)}\right\}\right)-1+
$$
The residual terms are $\propto (\delta a)^2$:
$$
+\left\{\left[(p_i(\xi_1)\cdot p_f(\xi_1))-1\right]
(l^*\cdot k)(l\cdot k)
-(l^*\cdot k)(l\cdot p_i(\xi_1))(k\cdot p_f)-
(l^*\cdot k)(l\cdot p_f(\xi_1))(k\cdot p_i)\right\}\times
$$
$$
\times\frac{(a(\xi)-a(\xi_1))^2}{2(k\cdot p_i)(k\cdot p_f)}=
$$
Use Eq.(\ref{eq:cons}) and introduce wherever reasonable the functions of $\xi$:
$$
=(l^*\cdot p_{i}(\xi_1))\left(l\cdot p_f(\xi)\right)+(l^*\cdot p_{f}(\xi_1))\left(l\cdot p_i(\xi)\right)+
$$
$$
+(l^*\cdot p_{i}(\xi_1))\left(l\cdot k\right)\frac{\left[a(\xi)-a(\xi_1)\right]^2}{2(k\cdot p_f)}+
(l^*\cdot p_{f}(\xi_1))\left(l\cdot k\right)\frac{\left[a(\xi)-a(\xi_1)\right]^2}{2(k\cdot p_i)}+
$$
$$
+\left\{\left[(p_i(\xi_1)\cdot p_f(\xi_1))-1\right]
(l^*\cdot k)(l\cdot k)
-(l^*\cdot k)(l\cdot p_i(\xi_1))(k\cdot p_f)-
(l^*\cdot k)(l\cdot p_f(\xi_1))(k\cdot p_i)\right\}\times
$$
$$
\times\frac{(a(\xi)-a(\xi_1))^2}{2(k\cdot p_i)(k\cdot p_f)}+
$$
The total of all terms proportional to $(a(\xi)-a(\xi_1))^2$ given above 
vanishes. This can be shown using the following properties: the tensor 
$l\otimes l^*$ is Hermitian; the difference $p_f-p_i$ is the total of two 
terms, proportional to $k$ and $k^\prime$, the latter being orthogonal to $l$. 
The residual terms are:
$$
+\frac12\left(p_{i}(\xi_1)\cdot\left\{p_f(\xi)+k\frac{\left[a(\xi)-a(\xi_1)\right]^2}{2(k\cdot p_f)}\right\}\right)+
$$
$$
+\frac12\left(p_{f}(\xi_1)\cdot\left\{p_i(\xi)+k\frac{\left[a(\xi)-a(\xi_1)\right]^2}{2(k\cdot p_i)}\right\}\right)-
$$
The last term, $-1$, we transform using Eq.(\ref{eq:pp})
$$
-\frac{1}2 \left[a(\xi)-a(\xi_1)\right]^2-\frac12(p_i(\xi)\cdot p_i(\xi_1))-
\frac12(p_f(\xi)\cdot p_f(\xi_1))=
$$
$$
=(l^*\cdot p_{i}(\xi_1))\left(l\cdot p_f(\xi)\right)+(l^*\cdot p_{f}(\xi_1))\left(l\cdot p_i(\xi)\right)-\left(\left[p_f(\xi)-p_i(\xi)\right]\cdot\left[p_f(\xi_1)-p_i(\xi_1)\right]\right)+
$$
$$
+\frac{\left[a(\xi)-a(\xi_1)\right]^2\left((k\cdot p_i)-(k\cdot p_f)\right)^2}{4
(k\cdot p_i)(k\cdot p_f)}.
$$
On substituting this into the integral expression for the matrix
element, some terms give zero contributions to the integral. 
Particularly, the following integrals vanish, as long as the
expressions in the square brackets are the perfect time derivatives:
$$
\int{\left[(k^\prime\cdot p_{i,f}(\xi))\exp(\int^\xi{\frac{(k^\prime\cdot p_{i,f}(\xi_1))}{(k\cdot p_{f,i})}d\xi_2})\right]d\xi}=0.
$$
From here it is also easy to derive that:
$$
\int{\left[(p_i(\xi)-p_f(\xi)-k^\prime)\exp(\int^\xi{\frac{(k^\prime\cdot p_{i,f}(\xi_1))}{(k\cdot p_{f,i})}d\xi_2})\right]d\xi}=0.
$$
Therefore, with the transformed product of the Dirac matrices, which
should be multiplied by a factor of two (as long as we should sum over
the final electron polarization instead of averaging), we obtain: 
\begin{equation}\label{eq:dWfi}
\frac{dW_{fi}}{d(k\cdot k^\prime)d^2{\bf k}^\prime_\perp}=
\frac{\alpha \int_{\xi_-}^{\xi_+} {\int_{\xi_-}^{\xi_+}{T(\xi)T(-\xi_1)]D
d\xi d\xi_1}}}{(2\pi\lambdabar_C)^2
(k\cdot k^\prime)(k\cdot p_i)(k\cdot p_f)},
\end{equation}
where
$$
D=(l^*\cdot p_{i}(\xi_1)) (l\cdot p_i(\xi))+
\frac{\left[a(\xi)-a(\xi_1)\right]^2\left((k\cdot p_i)-(k\cdot p_f)\right)^2
}{8 (k\cdot p_i)(k\cdot p_f)}.
$$

The matrix element may 
also
be 
summed, if desired, for two possible directions of the polarization
vector. The second term in the integrand is simply multiplied by two, while in 
the first one 
the negative of the metric tensor 
 should be substituted for
the product of the polarization vectors 
(see 
Section 8
in \cite{lp}), so that  
$-\left(p_i(\xi)\cdot p_i(\xi_1)\right)$
substitutes for 
$\sum_l{(l^*\cdot p_{i}(\xi_1)) (l\cdot p_i(\xi))}$. 
The latter may be transformed using
Eq.(\ref{eq:pp}), thus,  giving:
$$
\sum_{l}{D}=-1+
\frac{\left[a(\xi)-a(\xi_1)\right]^2\left((k\cdot p_i)^2+(k\cdot p_f)^2\right)
}{4 (k\cdot p_i)(k\cdot p_f)}.
$$ 
\subsection{Calculation of the matrix element: integration}
Now perform integration over ${\bf k}_\perp$.  
On developing the dot-product, $(k^\prime\cdot p_i)$, in 
$T(\xi)$ in the TST  metric, $G^{\mu\nu}$, one can find: 
$$
T(\xi)T(-\xi_1)=\exp\left[i(T_1+T_2)\right], 
$$
where 
$$T_1=\frac{(k\cdot p_i)}{2\lambdabar_C(k\cdot k^\prime)(k\cdot p_f)}\left(\frac{(k\cdot k^\prime)}{(k\cdot p_i)}\left<{\bf p}_{\perp i}\right>-
{\bf k}^\prime_{\perp}\right)^2(\xi-\xi_1),
$$
$$
T_2=\frac{ (k\cdot k^\prime) \left\{(\xi-\xi_1)+\int_{\xi_1}^\xi{
\left[{\bf a}(\xi_2)-\left<{\bf
    a}\right>\right]^2d\xi_2}\right\}}{2\lambdabar_C(k\cdot p_i)(k\cdot p_f)}, 
$$
$$
\left<{\bf a}\right>=\frac{\int_{\xi_1}^\xi{{\bf
      a}d\xi_2}}{\xi-\xi_1},\qquad \left<{\bf p}_{\perp i}\right>={\bf
  p}_{\perp 0 i}-\left<{\bf a}\right>.$$ 
Integration over $d^2{\bf
  k}^\prime_\perp$ 
is performed using the following formula: 
$$
\int{\exp\left(iT_1\right)d^2{\bf k}_\perp}=
\pi\int_0^\infty{\exp\left(iT_1\right)d\left(\frac{(k\cdot k^\prime)}{(k\cdot p_i)}\left<{\bf p}_{\perp i}\right>-
{\bf k}^\prime_{\perp}\right)^2}.
$$
The right hand side is proportional to $\exp(iT_1)_{|{\bf
    k}_\perp|\rightarrow\infty}-1$, where the rapidly oscillating at
large $|{\bf k}_\perp|$ term
results in a vanishing contribution to the integral over $d\xi d\xi_1$
and should be neglected. The integrated over ${\bf k}_\perp$ 
probability is: 
$$
\frac{dW_{fi}(\xi_-,\xi_+)}{d(k\cdot k^\prime)}=
\frac{\alpha \int_{\xi_-}^{\xi_+}
  {\int_{\xi_-}^{\xi_+}{\frac{i\exp(iT_2)}{\xi-\xi_1}\sum_{l}{D(\xi,\xi_1)}
d\xi d\xi_1}}}{2\pi\lambdabar_C(k\cdot p_i)^2}.
$$ 
In  
strong fields 
the following estimates may be applied:   
$$
(k\cdot k^\prime)\sim \lambdabar_C(k\cdot p_i)^2\left|\frac{d{\bf a}}{d\xi}\right|,\qquad
dW_{fi}\sim\alpha\left|(\xi_+ - \xi_-)\frac{d{\bf a}}{d\xi}\right|.$$ 
Now the bounds for $\xi_+-\xi_-$ can be {\it consistently} introduced: 
\begin{equation}\label{consistency}
\left|d{\bf a}/d\xi\right|^{-1}\ll\xi_+-\xi_-\ll
\min\left(\alpha^{-1}\left|d{\bf a}/d\xi\right|^{-1},1\right).
\end{equation}
Under these bounds, first, the time interval (\ref{eq:time}) is much
greater than the {\it formation time}. Therefore, once the double integral
over $d\xi d\xi_1$ is transformed into the integral over 
$d(\xi-\xi_1)d(\xi+\xi_1)/2$, the span to integrate over 
$d\theta=d(\xi-\xi_1)$
can be extended towards $\pm\infty$. Hence, the emission probability 
becomes linear in $\xi_+-\xi_-$, because only the integration over 
$d(\xi+\xi_1)/2$ is performed from $\xi_-$ till $\xi_+$: 
$$dW_{fi}(\xi_-,\xi_+)=(dW/d\xi)(\xi_+-\xi_-).$$ 
On the other hand  the difference, $\xi_+-\xi_-$, is to be small enough, so
that the probability of emission within the time interval 
of Eq.(\ref{eq:time}) is much less (or at least less) 
than unity:
\begin{equation}\label{eq:upperbound}
\int{\frac{dW_{fi}}{d{\bf k}^\prime_\perp d(k^\prime\cdot k)}d{\bf k}^\prime_\perp d(k^\prime\cdot k)}\ll1.  
\end{equation}
Therefore,  
perturbation theory is applicable. In addition, 
the emission probability can be expressed in terms of the local 
electric field. Note, that consistency in
(\ref{consistency}) is ensured in relativistically strong
electromagnetic fields as long as $\alpha\ll1$, with no restriction on
the magnitude of the electromagnetic field experienced by an electron.

Under the condition (\ref{consistency}) 
in the integral
  over $d\theta$ the small differences in the vector potential may be
  linearized: 
$$
-[
a(\xi)-a(\xi_1)
]^2= [
{\bf a}(\xi)-{\bf a}(\xi_1)
]^2=
\left(
\frac{d{\bf a}}{d\xi}\right)^2\theta^2,
$$
$$
T_2=\frac
{ (k\cdot k^\prime)
  \left\{
\theta+\frac{\theta^3}{12}
\left(\frac{d{\bf a}}{d\xi}\right)^2\right\}}{2\lambdabar_C(k\cdot
  p_i)(k\cdot p_f)},
\qquad -\sum_{l}{D(\xi,\xi_1)}=1+\left(
\frac{d{\bf a}}{d\xi}\right)^2\theta^2\frac{(k\cdot p_i)^2+(k\cdot p_f)^2
}{4 (k\cdot p_i)(k\cdot p_f)}.
$$
After this the integral,
$$
\frac{dW}{d\xi d(k\cdot k^\prime)}=-\frac{\alpha \int_{-\infty}^{\infty}
  {\frac{\sin(T_2)}{\theta}\sum_{l}{D(\xi,\xi_1)}
d\theta}}{2\pi\lambdabar_C(k\cdot p_i)^2},
$$
by means of a substitution, $z=|d{\bf a}/d\xi|\theta/2$, may be expressed in terms of the MacDonald functions,
using the following equations:
$$
\int\limits_{-\infty}^{+\infty}{\cos\left[\frac32r\left(\frac{z^3}3+z\right)\right]dz}=\frac2{\sqrt{3}}K_{1/3}(r),
$$
(see Eq.(8.433) in \cite{gr})
$$
\int\limits_{-\infty}^{+\infty}{z\sin\left[\frac32r\left(\frac{z^3}3+z\right)\right]dz}=\frac2{\sqrt{3}}K_{2/3}(r),
$$
$$
\int\limits_{-\infty}^{+\infty}{\frac1z\sin\left[\frac32r\left(\frac{z^3}3+z\right)\right]dz}=-\frac2{\sqrt{3}}
\int\limits_r^{+\infty}{K_{1/3}(r^\prime)dr^\prime},
$$
as well as the recurrent relationships between the MacDonald functions:
$$
2\frac{dK_\nu(r)}{dr}=-K_{\nu-1}(r)-K_{\nu+1}(r), \qquad
K_{\nu-1}(r)-K_{\nu+1}(r)=-\frac\nu rK_\nu(r),
$$
and the identity, $K_{-\nu}(r)=K_\nu(r)$.  In this way we arrive at
Eq.(\ref{eq:probabf}).
The advantage of the MacDonald functions is the simplicity and fast
convergence of the integral representation for them:
$$
K_{\nu} (r)=
\int_0^{\infty}{ \exp\left[-r \cosh(z)\right]\cosh(\nu
    z) dz }, 
$$
$$\int_r^\infty{K_{\nu} (r^\prime)dr^\prime}=
\int_0^{\infty}{ \exp\left[-r \cosh(z)\right]\frac{\cosh(\nu
    z)}{\cosh(
    z)} dz },
$$
(see Eq.(8.432) in \cite{gr}). In numerical simulations, therefore, the
MacDonald functions are very easy to use.
\section{Appendix B. Photon in the QED-strong field: absorption
  probability}
Here we consider a photon with the wave four-vector $k^\prime$ and find
a probability of its absorption in the strong 1D wave field with
producing electron and positron, their four-momenta being $p_e$ and
$p_p$ correspondingly. 

The matrix element for the absorption probability is now related to the
element of the electron phase volume. The wave function for the
absorbed photon is the complex conjugated wave function of the emitted
photon. The bi-spinor amplitude for positron is the Dirac conjugated
amplitude for electron. With these minor changes, the matrix element for
the absorption probability is transformed as follows: 
$$
\frac{dW_{fi}(\xi_-,\xi_+)}{d(k\cdot p_e)d^2{\bf p}_{e\perp}}=
\frac{\alpha \left|
   \int_{\xi_-}^{\xi_+}{T(\xi)\hat{u}(p_e)(\gamma\cdot l)u(p_p)d\xi
       }\right|^2}{(4\pi\lambdabar_C)^2(k\cdot k^\prime)(k\cdot
  p_e)(k\cdot p_p)},
$$
where 
$$
T(\xi)=
\exp\left[\frac{i\int^\xi{(k^\prime\cdot p_{e,p}(\xi_2))d\xi_2}}{\lambdabar_C(k\cdot p_{p,e})}
\right],
$$ 
and 
\begin{equation}\label{eq:cons1}
p^\mu_e(\xi)+p^\mu_p(\xi)=(k^\prime)^\mu+\frac{(k^\prime\cdot p_{e,p}(\xi))}
{(k\cdot k^\prime)-(k\cdot p_{e,p})}k^\mu . 
\end{equation}
Eq.(\ref{eq:cons1}) may be obtained from Eq.(\ref{eq:cons}) as well as the new
phase multiplier may be obtained from the earlier derived phase
multiplier by virtue of the following transformation:
\begin{equation}\label{crossinv}
p_f\rightarrow p_e,\qquad p_i\rightarrow -p_p,\qquad
k^\prime\rightarrow -k^\prime,
\end{equation}
because the photon emission and the photon absorption are two
cross-invariant channels of the same reaction. Then,
to calculate the matrix element, one can re-write it as the 
double integral over $d\xi d\xi_1$. The integrand in this
double integral includes the phase multiplier and the following
matrix product: 
\begin{equation}\label{eq:matproduct1}
\hat{u}(p_e(\xi))(\gamma\cdot l)u(p_p(\xi))
       \left[\hat{u}(p_e(\xi_1))(\gamma\cdot
         l)u(p_p(\xi_1))\right]^*=...
\end{equation} 
Now we should follow the way we used to expand Eq.(\ref{eq:matproduct})
with the following modifications. The polarization matrix for the positron is $1/2[(\gamma\cdot
  p_p(\xi))-1]$. While applying Eq.(\ref{eq:QEDpropagator}) to a positron, the
dimensionless vector potential, which had been defined above in terms
of a negative charge of electron, should now be taken with the opposite
sign, therefore: 
$$
...=\frac14{\rm Sp}\{
\left[1+\frac{(\gamma\cdot k)}{2(k\cdot p_e)}
(\gamma\cdot[ a(\xi_1)-
a(\xi)])\right]\times
$$
$$
\times
\left[(\gamma\cdot p_e(\xi))+1\right](\gamma\cdot l)\times\nonumber\\
\left[1-\frac{(\gamma\cdot k)}{2(k\cdot p_p)}
(\gamma\cdot [a(\xi)-a(\xi_1)])\right]
\left[(\gamma\cdot p_p(\xi_1))-1\right](\gamma\cdot l^*)    
\}.
$$  
We see that again the transformation Eq.(\ref{crossinv}) allows us to
obtain the above matrix product from that one we derived in developing
Eq.(\ref{eq:matproduct}) in Appendix A. However, in addition to the
modifications listed in Eq.(\ref{crossinv}) we need to transform
$l\rightarrow l^*$ and change the whole sign of the matrix product. Now
we can apply this transformation procedure to the resulting expression
for the emission probability, Eq.(\ref{eq:dWfi}), and obtain the
result for the absorption probability: 
\begin{equation}\label{eq:dWfi1}
\frac{dW_{fi}}{d(k\cdot p_e)d^2{\bf p}_{e\perp}}=
\frac{\alpha \int_{\xi_-}^{\xi_+} {\int_{\xi_-}^{\xi_+}{T(\xi)T(-\xi_1)]D
d\xi d\xi_1}}}{(2\pi\lambdabar_C)^2
(k\cdot k^\prime)(k\cdot p_e)(k\cdot p_p)},
\end{equation}
where
$$
D=-(l^*\cdot p_{e}(\xi_1)) (l\cdot p_e(\xi))+
\frac{\left[a(\xi)-a(\xi_1)\right]^2(k\cdot k^\prime)^2
}{8 (k\cdot p_p)(k\cdot p_e)}.
$$

Note an interesting polarization property: in the field of linearly
polarized 1D wave the emission probability is {\it maximal} for photons with
the same polarization as that for 1D wave, but the absorption
probability for such photons is {\it minimal} (the
polarization-dependent term in $D$ is negative). We still employ here
the probabilities averaged over the photon polarizations, but we think
this approach may be accurate only when applied to the processes in
circularly polarized strong wave fields.   

On integrating the averaged absorption probability over $d{\bf p}_{e\perp}$, we obtain the following
expression:
\begin{equation}\label{eq:probabf1}
\frac{dW_{fi}}{d(k\cdot k^\prime)d\xi}=
\frac{\alpha\left(-\int_{r}^\infty{K_{5/3}(y)dy}+\frac{(k\cdot
    k^\prime)^2}{(k\cdot p_e)(k\cdot p_p)}
K_{2/3}(r)\right)}{\sqrt{3}\pi\lambdabar_C(k\cdot k^\prime)^2},
\end{equation}
$$ 
r=\frac{(k\cdot
  k^\prime)}{\chi_e(k\cdot p_p)},\quad
\chi_e=\frac32(k\cdot p_e)\left|\frac{d{\bf a}}{d\xi}\right|\lambdabar_C,
$$
which is used as the kernel in the collision integral in the present
paper.

\end{document}